\documentclass[conference]{IEEEtran}
\IEEEoverridecommandlockouts
\usepackage{cite}
\usepackage{amsmath,amssymb,amsfonts}
\usepackage{algorithmic}
\usepackage{graphicx}
\usepackage{textcomp}
\usepackage{xcolor}
\usepackage{subcaption}
\usepackage{caption}
\usepackage{dblfloatfix}
\def\BibTeX{{\rm B\kern-.05em{\sc i\kern-.025em b}\kern-.08em
    T\kern-.1667em\lower.7ex\hbox{E}\kern-.125emX}}

\begin{document}

\title{QuAS: Quantum Application Score for benchmarking the utility of quantum computers}

 \author{\IEEEauthorblockN{Koen J. Mesman}
 \IEEEauthorblockA{\textit{QAIMS} \\
 \textit{Delft University of Technology}\\
 Delft, Netherlands \\
 k.j.mesman@tudelft.nl}
 \and
 \IEEEauthorblockN{Ward van der Schoot}
 \IEEEauthorblockA{\textit{Applied Cryptography and Quantum Algorithms} \\
 \textit{TNO}\\
 The Hague, Netherlands\\
 ward.vanderschoot@tno.nl}
 \and
 \IEEEauthorblockN{Matthias M\"oller}
 \IEEEauthorblockA{\textit{Applied Mathematics} \\
 \textit{Delft University of Technology}\\
 Delft, Netherlands \\
 m.moller@tudelft.nl}
 \and
 \IEEEauthorblockN{Niels M. P. Neumann}
 \IEEEauthorblockA{\textit{Applied Cryptography and Quantum Algorithms} \\
 \textit{TNO}\\
 The Hague, Netherlands\\
 niels.neumann@tno.nl}
 }

\maketitle

\begin{abstract}
Benchmarking quantum computers helps to quantify them and bringing the technology to the market. Various application-level metrics exist to benchmark a quantum device at an application level. This paper presents a revised holistic scoring method called the Quantum Application Score (QuAS) incorporating strong points of previous metrics, such as QPack and the Q-score. We discuss how to integrate both and thereby obtain an application-level metric that better quantifies the practical utility of quantum computers. We evaluate the new metric on different hardware platforms such as D-Wave and  IBM as well as quantum simulators of Quantum Inspire and Rigetti.
\end{abstract}


\section{Introduction}
With the development of quantum computers, interest in the field is rising quickly and broadens to an increasing number of (sub)fields. 
On the one hand, end users start to realize that these devices could significantly speed up their business processes, while investors learn that timely investments into quantum computing could yield them large profits. 
However, quantum computing is a completely novel field for both groups. 
Current work is still very technical, making it hard for these new parties to gauge the current and future applicability of quantum devices. 

At the same time, research organizations are developing new quantum devices at a rapid rate.
Currently, many different technologies are utilized to realize a quantum device, examples include trapped ions~\cite{Noel_2022}, atoms~\cite{Saffman_2016,Beterov_2018}, spin systems~\cite{Struck2016}, and superconducting circuits~\cite{annurev_soa,orlando_flux}.
At the time of writing, more than 100 different quantum devices have been developed, each with their own strengths and weaknesses. This zoo of quantum devices is making it even harder for new quantum-interested parties to gauge the applicability of these devices.

The solution to this problem is given by so-called \textit{quantum metrics}.
Quantum metrics are ways of measuring and hence comparing the performance of quantum devices. This can be done in a variety of ways. Firstly, there are a variety of aspects that can be measured, such as size, accuracy, runtime, or energy usage. Secondly, there are different levels at which the performance can be evaluated. 
Van der Schoot et al. distinguish between three different levels of quantum metrics in~\cite{vanderSchoot2023}, namely component-level metrics, system-level metrics, and application-level metrics. 
Component-level metrics focus on evaluating specific components of devices, such as the decoherence time of qubits, and the fidelity of quantum gates. System-level metrics gauge the performance of a quantum device as a whole, for example through the quantum volume~\cite{Cross:2019} or mirror circuits~\cite{Proctor2021}. Application-level metrics take a more heuristic approach and consider only the ability of quantum devices to solve certain problems.

This final level of metrics is vital for the uptake of quantum computing by industry. 
By focusing on the application, these metrics are understandable even for investors or end users less familiar with quantum computing. In addition, with these metrics, quantum providers can see what they should focus on to best serve their current or future customers.

This work hence focuses on application-level metrics and proposes a new application-level metric called the Quantum Application Score (QuAS). While there are already a variety of available application-level metrics, we believe that there are still improvements to be made. By considering the state of the art of application-level metrics, we combine their strengths and avoid their weaknesses to come up with a metric which is perfectly suitable for gauging the performance of quantum devices on an application level. Further discussion as to how our metric relates to the state of the art and how we expand upon it can be found in the related work chapter.

This work will be structured as follows. 
Firstly, Section~\ref{sec:related_work} explains prior work in application-level metrics and why our metric solves existing issues with other metrics. 
Section~\ref{sec:method} discusses the newly proposed metric called the QuAS. 
The results after evaluating this metric on different hardware platforms are given in Section~\ref{sec:results}.
This work ends with a conclusion and discussion in Section~\ref{sec:conclusion}. 

\section{Related work}
\label{sec:related_work}
\subsection{The field of quantum metrics}
With the start of the production of quantum computers, quantum computer capabilities were measured using component-level metrics such as the number of qubits, fidelity of gates and measurement operators, and decoherence times such as $T_1$ and $T_2$ times~\cite{Youssef:2020}. In this way, researchers and manufacturers could see on a low level at which parts their device was excelling, and which parts required more attention.

 When quantum devices started to run actual circuits, it became clear that the capability to run circuits could not be fully captured by these component-level metrics. That is why system-level metrics were introduced, starting with the Quantum Volume, which measures the `size' of the largest representative quantum circuit a processor can faithfully execute~\cite{Cross:2019}. Soon after, other ways of measuring performance at implementing circuits were introduced, such as mirror circuits~\cite{Proctor2021} and Circuit Layer Operations Per Seconds~\cite{CLOPS}.

More recently with the first implementations of toy-level problems, researchers realized that while circuit implementations form an important part of running quantum algorithms, they do not capture the full picture. For this reason, metrics at an even higher level were considered, later named application-level metrics.

Currently, already a variety of application-level metrics exists, widely varying in their approach and focus. In this overview, we highlight a few metrics in chronological order. For each metric, at least the considered problem and the measured quantity are given. It should be noted that this overview is incomplete.

In 2020, Dong and Lin suggested the Quantum LINPACK as an equivalent to the LINPACK metric used to benchmark classical supercomputers. Just like LINPACK is measured by a classical device's performance at solving linear equations with random dense matrices, Quantum LINPACK measures a quantum processor's performance at the Quantum Linear Solving Problem with random block-encoding matrices~\cite{DongLin:2021}. The metric outcome is the probability of success in finding the correct output state.

In March 2021, Mesman et al. introduced the QPack metric~\cite{mesman2021qpack, donkers2022qpack}. This benchmark offers a generic framework, in which many problems can be used to benchmark a quantum processor. Examples are the Max-Cut problem (MCP) and the Dominating Set Problem. Given a certain problem and quantum device, the device is used to solve the problem, from which the runtime, accuracy, scalability and capacity of the device are computed. The runtime is computed as the number of gates per second; the accuracy is computed as how close the quantum solution gets to the classical-best solution; the scalability is computed as the exponent in the fit between the runtime and the problem size; and the capacity is computed as the largest problem size for which the quantum solution lies within a certain bound of the classical-best solution. By plotting these obtained four values in a radar plot, the QPack of the device can be computed as the area of the figure spanned by these values in the radar plot.

In July 2021, Martiel et al. proposed the Q-score metric~\cite{Martiel:2021}. This was originally designed to benchmark the performance of a gate-based quantum processor at solving the MCP. It is defined as the largest problem instance for which the processor significantly outperforms a random approach. Later, Van der Schoot et al. showed how the metric could easily be extended to consider other problems as well, such as the Max Clique problem~\cite{vanderSchoot2023}. In addition, they showed that the metric is also naturally suitable to benchmark other quantum processors, such as quantum annealers and photonic devices, as well as classical solvers.

In October 2021, Lubinski et al. introduced the QED-C benchmarking suite~\cite{Lubinski:2021}. Their idea was that a single problem is insufficient to encapsulate the performance of a quantum processor at an application-level. That is why they propose to compare quantum devices by using a suite of representative problems. This suite contains toy problems, such as Deutsch-Jozsa, quantum subroutines, such as the Quantum Fourier Transform, full quantum algorithms, such as Shor's algorithm, and classical problems, such as the Max-Cut problem. For each of the problems, the fidelity is computed for a variety of circuit sizes, and plotted in volumetric space depicting width and depth. The fidelity is then represented using colors. Finally, the outputted QED-C metric of a quantum device is given by these diagram(s) for the various problems.

In February 2022, Tomesh et al. developed the SupermarQ benchmarking suite~\cite{Tomesh:2022}. The suite concerns a variety of problems to gauge the performance of a quantum processor, such as $GHZ$-state generation, error-correcting code routines, the Quantum Approximate Optimization Algorithm (QAOA), and the Variational Quantum Eigensolver (VQE). For each problem, the performance metric is defined differently and scaled to the $[0,1]$-interval. Benchmarking a quantum processor means measuring these performance scores for different problems. The final SupermarQ score of the processor follows by plotting the respective subscores in a space of dimension equal to the number of different problems and equals the hypervolume of the shape spanned by the performance scores. 

From the above, it can be seen that there exists a wide variety of application-level metrics. While some metrics specifically focus on the performance of specific problems, such as Quantum LINPACK, other metrics evaluate quantum devices given a suite of problems, such as the QED-C suite, QPack, and SupermarQ. In addition, different quantities are used to measure performance, such as fidelity, problem size, accuracy, and runtime. Lastly, the output from a given metric can vary widely, from a single number to multiple diagrams.

\subsection{Why do we need yet another application-level metric?}
In this work, we establish a new application-oriented quantum metrics called QuAS. Even with the apparent abundance of application-oriented quantum metrics in literature, we believe none of them meets user demands placed on metrics:
Specifically, we believe that a quantum metric should meet the following requirements, based on earlier requirements by the Boston Consultancy Group and Atos~\cite{BCG, Martiel:2021}. A quantum metric should be
\begin{enumerate}
    \item easy to \textbf{compare}, preferably consisting of just a single number, while also having the possibility to consider subparts of the metric. 
    \item \textbf{applicable} to any application or problem area. This is most naturally achieved by defining the metric as a framework or as a suite of metrics. 
    \item \textbf{scalable} for both noisy and fault-tolerant quantum hardware. This specifically means that metrics that require classical simulations of many qubits are inherently impossible.
    \item \textbf{hardware-agnostic}, both on a technology-level and a paradigm-level. The latter means that the metric should not favor one quantum paradigm over the other, such as quantum annealing over gate-based quantum computing. The former means that the metric should not favor certain technologies within a quantum paradigm, such as spin qubits over neutral atom qubits.
    \item an \textbf{equitable analysis} between different key performance indicators (KPIs) relevant to the application, such as runtime, accuracy, and problem size. In addition, the relative importance of the different KPIs should be flexible in the definition of the metric.
\end{enumerate}

None of the current application-level quantum metrics meet all five requirements. Still, various metrics exist that meet some of the requirements, such as the QED-C metrics meeting requirements 2, 3, and 4, and the Q-score meeting requirements 1, 2, 3, and 4. The QuAS defined in this work is the first metric to meet all requirements, thereby justifying the QuAS metric.

\section{Method}
\label{sec:method}
In this section, we explicitly define the QuAS metric. 
The QuAS metric is a framework to benchmark a quantum device with a couple of degrees of freedom that a user can choose depending on the intended application.
The degrees of freedom are the problem instance(s), the KPIs, and the weight of the different KPIs.
In this work, the QuAS metric is defined using three specific KPIs (size, accuracy, and runtime) with equal weight, and three different example problems (Max-Cut, Ising, and Travelling Salesman Problem). 
However, it is important to mention that all these degrees of freedom can be adapted to the preference of the end user.

This section will be structured as follows: First, the chosen KPIs and example problems will be discussed; Afterward, the intuition behind the QuAS metric will be explained; and finally, the precise definition of QuAS will be given.


\subsection{KPIs}
An important characteristic of the QuAS metric is that it can be defined using any set of KPIs. To show the workings of the metric, the QuAS metric will be implemented using three example KPIs. Based on earlier work~\cite{donkers2022qpack, Schoot2022EvaluatingQscore}, the KPIs \textit{problem size}, \textit{accuracy} and \textit{runtime} have been chosen. Naturally, these three KPIs intricately depend on one another: for example, if a solver is given more time, it can achieve a higher accuracy. The metric should take these dependencies into account.

\subsubsection{Problem Size}
The size of a problem that can be solved with the quantum computer. This relates directly to the number of available qubits.

\subsubsection{Accuracy}
The accuracy is a measure of the quality of the solution. For a given problem size, different solvers will find a certain solution. The accuracy measure is then determined by comparing this solution $S$ to a solution $S_{heur}$ found by a classical (heuristic) solver.
Explicitly, the accuracy KPI is defined as $$1-\frac{S_{heur}-S}{S_{heur}}.$$

The accuracy score will therefore always be relative to a benchmark classical algorithm. This has advantages and disadvantages. The main advantage is that the resulting metric is scalable, which is not the case if for example a brute-force optimal solution is used as comparison. Alternatively, a current best classical solver could be chosen, but as in time better algorithms are found, the baseline would need to be constantly updated, making it impractical. The drawback is that the obtained accuracy is a weak score: an obtained quantum solution can be several factors better compared to the heuristic, while still underperforming compared to exact algorithms, or even computationally comparable algorithms. This disadvantage is acceptable, as the accuracy is defined inherently relative: every solver that gets benchmarked using this metric, has the accuracy defined using the same heuristic score. 

It is good to emphasize that this could result in an accuracy larger than 1, but that is not an issue in itself. Specifically, this does not imply a hard quantum advantage, but only a potential advantage over the used classical heuristic, depending on the used runtimes. 

\subsubsection{Runtime} 

Fundamental to solving computational problems is the total runtime of solving the problem. There are several ways in which the runtime can be defined. For practical use, the end-to-end solution time is the only relevant time. For NISQ systems, the actual quantum (QPU) time is more relevant, as current remote systems can suffer from queue times. In this work, the wall-time is considered, which is defined as the time taken for the classical optimizer and quantum processor to converge to a solution.

\subsection{Example Problems}
Another important characteristic of the QuAS metric is that it is defined in a problem-agnostic way. Because of this, the metric can be used to benchmark a (quantum) device at any problem of interest, as long as the chosen KPIs can be consistently measured using a single-number value. To showcase its applicability, the metric will be applied using three different problems often seen in quantum computing, namely the Max-Cut Problem, the Ising model, and the Travelling Salesman Problem.

\subsubsection{Max-Cut Problem}
The Max-Cut problem is a famous problem in graph theory. The problem takes as input a (potentially weighted) undirected graph, and seeks to find a partition of the vertex set into two, so that the total weight of edges between the two sets is maximized. Such a partition is called a \textit{cut}, and the total weight of these edges is called the \textit{cost} of the cut. The Max-Cut problem is hence the problem of finding the cut which has a maximal cost. In this work, the Max-Cut problem is considered for a family of Erd\"os-R\'enyi $(N,\frac12)$-graphs: random graphs of $N$ vertices, and each edge is added with probability $\frac12$. The general Max-Cut problem, as well as the problem for these specific graphs, is NP-complete.

\subsubsection{Ising Model}
The Ising model, originating from physics, contains discrete variables $\sigma_i$ representing the magnetic dipole moment of atomic spins, which can equal $-1$ or $+1$. The aim of the problem is to minimise a certain Hamiltonian function of the form
\[
H = -\sum_{i,j}J_{ij}\sigma_i\sigma_j-\mu\sum_jh_j\sigma_j
\]
Various phenomena in physics act according to the Ising model, because of which it is often studied in quantum mechanics. In addition, it is easily shown that the Ising model is equivalent to the Quadratic Unconstrained Binary Optimization Problem (QUBO). As many problems in optimization can be written as QUBOs, these problems can also be written as an Ising model. The Ising model is known to be NP-complete.

\subsubsection{Travelling Salesman Problem}
The Travelling Salesman problem is a famous combinatorial problem which considers a network of cities, connected by roads of a certain length. The goal is to find a walk that visits each city exactly once and that has a minimal total length. The traveling salesman problem is known to be NP-hard.

\subsection{QuAS definition}

In this subsection, the technical definition of the QuAS metric will be given. 
While the definition of the metric itself is quite technical, the intuition behind it is rather straightforward and natural.
Because of this, the intuition behind the metric will be explained first.
Note that the definition of the metric will be given using the chosen KPIs, but works in exactly the same way if other KPIs are chosen, such as energy usage, or in particular if more or less KPIs are chosen.

Given the performance data on the chosen example problem, each data point will consist of a triple (accuracy, runtime, size). For each fixed size, the (accuracy, runtime) data points span a shape in 2D-space, for which the Pareto front can be constructed. To limit the impact of outliers, this shape is fitted to a suitable curve, which will be defined below. The QuAS metric is then defined as the volume of the shape bounded by this curve, after scaling the different KPIs suitably. In addition, the QuAS metric allows an evaluation of how the different KPIs impact the QuAS metric. This can be done by extracting relevant features of the surface related to the different KPIs. This will be explained in further detail below.

\begin{figure}
    \centering
    \includegraphics[width=0.8\linewidth]{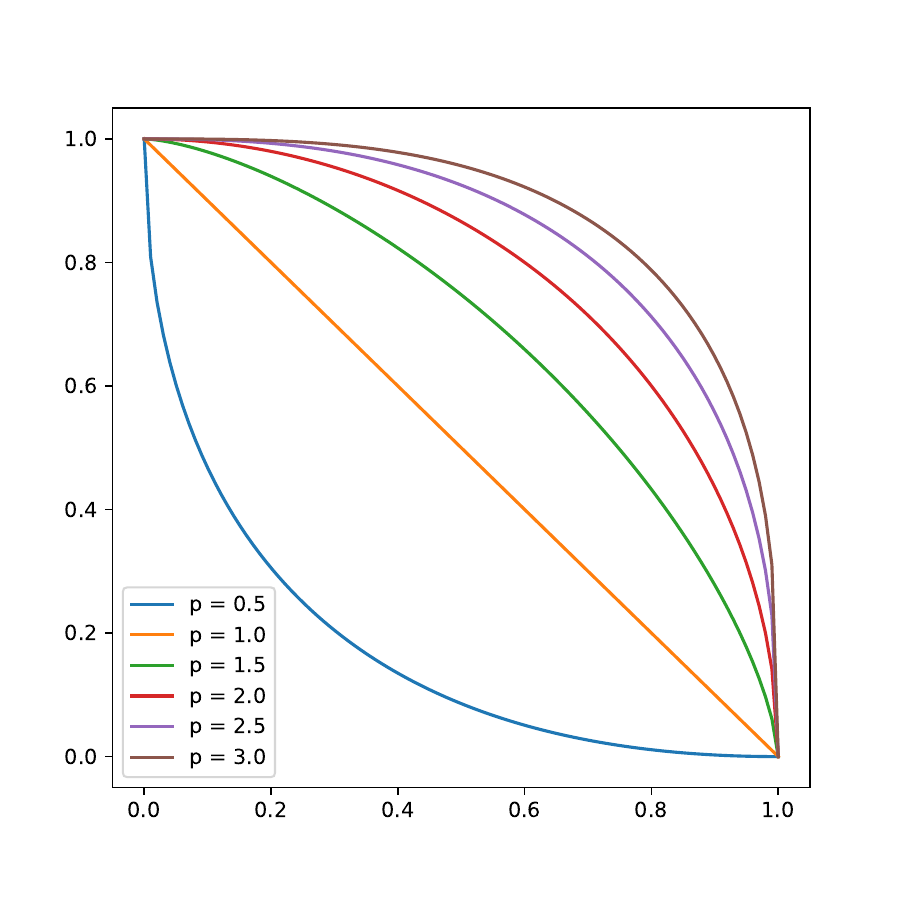}
    \caption{One quadrant of the unit circle for various values of $p$.}
    \label{fig:unitcircle}
\end{figure}

The distribution of the data is not known beforehand, therefore the shape of the curve depends significantly on the data. In this work, we have opted to fit a Lam\'e curve (ellipse in $L_p$ space) to the boundary of the cluster of data points. For the $n$ elements of a vector, the \textit{p}-norm is defined as:
\begin{equation}
    ||x||_p = \sqrt[\leftroot{-2}\uproot{2}p]{|x_1|^p + |x_2|^p + ... + |x_n|^p}
\end{equation}
The unit circle of this norm is given by $||x||_p = 1$.
Figure \ref{fig:unitcircle} shows a quadrant of the unit circle for different $p$-values.

The Lam\'e curve is an ellipsoid of this unit circle and is defined in the two-dimensional case as:
\begin{equation}
    |\tfrac{x}{a}|^p + |\tfrac{y}{b}|^p = 1,
\end{equation}
with scaling factors $a$ and $b$.

To fit the Lam\'e curve, the KPIs need to be normalized to the interval $[0,1]$. Therefore, each KPI is normalized by:
\begin{equation}
    x' = \frac{x-\text{min}(x)}{\text{max}(x)-\text{min}(x)},
\end{equation}
which results in offset $\alpha$:
\begin{equation}
    \alpha = \frac{\text{min}(x)}{\text{max}(x)-\text{min}(x)}.
\end{equation}
Note that this assumes that the different KPIs can be scaled so that they lie within a certain interval $[\text{min}(x), \text{max}(x)]$. This is true for the KPIs accuracy (by definition), and runtime.
The scaling factors $a$ and $b$ are then defined by $\text{max}(x)-\text{min}(x)$, one for each KPI. These scaling factors exactly detail the impact of the different KPIs, with a higher scaling factor indicating a larger impact on the QuAS metric.

The resulting cluster of data points given by the data points then lies within the unit-square $[0,1]^2$. 
Least Square Error (LSE) minimization using the Nelder-Mead optimizer then gives a best-fitted Lam\'e curve for these data points. 
This method results in two types of metrics: the QuAS metric, and a normalization factor for each KPI, except the problem size. 
\begin{figure}[ht]
    \centering
    \includegraphics[width=0.8\linewidth]{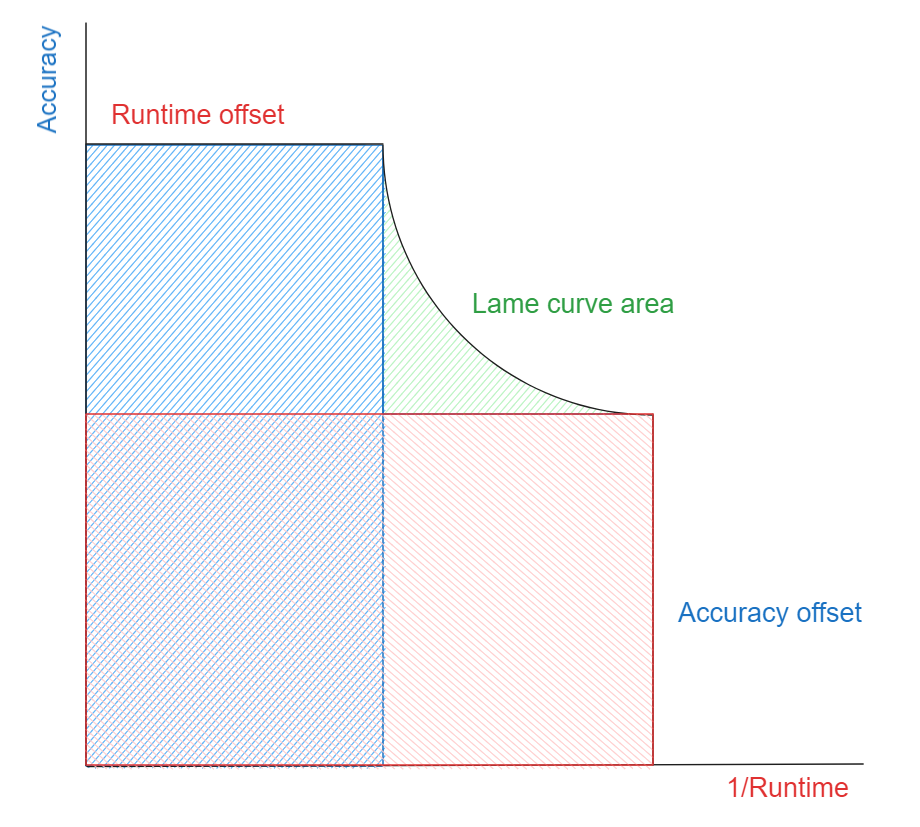}
    \caption{Elaboration of the area calculation used for the score. The green area of the Lam\'e curve is calculated according to equation \ref{eq:area}. The blue area is the runtime offset multiplied by maximum accuracy. The red area is the accuracy offset multiplied by the fastest runtime. The overlapping area then needs to be subtracted.}
    \label{fig:area}
\end{figure}
 With these factors, the area of the Lam\'e curve quadrant is defined as:
 \begin{equation}
 \label{eq:area}
     A = ab\frac{(\Gamma(1+\frac{1}{p}))^2}{\Gamma(1+\frac{2}{p})}
 \end{equation}
 where the gamma function ($\Gamma$) is defined as:
 \begin{equation}
     \Gamma(p) = \int_0^{\infty} t^{(p-1)}e^{-t} dt
 \end{equation}

\begin{figure*}[b]
    \centering
    \begin{subfigure}{.49\linewidth}
    \centering
        \includegraphics[width=0.7\linewidth]{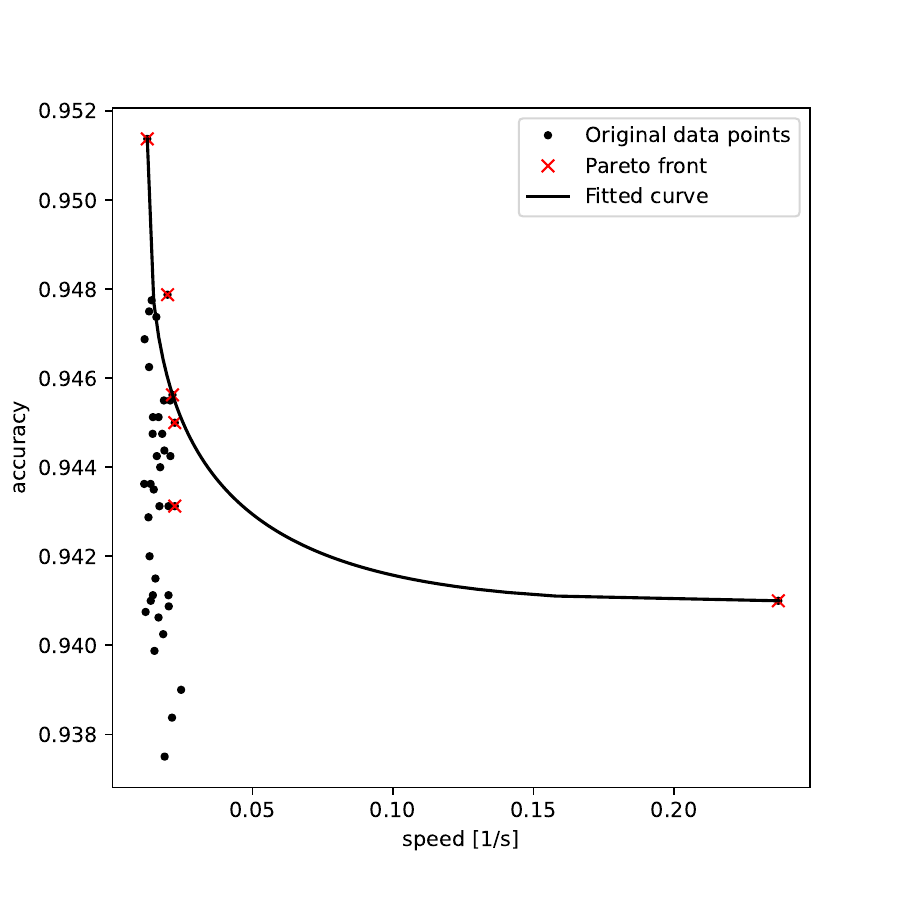}
        \caption{$L_p$ curve fitted to the Pareto front of the performance data. The tested problem size is 4.}
        \label{fig:normfit}
    \end{subfigure}
    \begin{subfigure}{.49\linewidth}
    \centering
        \includegraphics[width=0.7\linewidth]{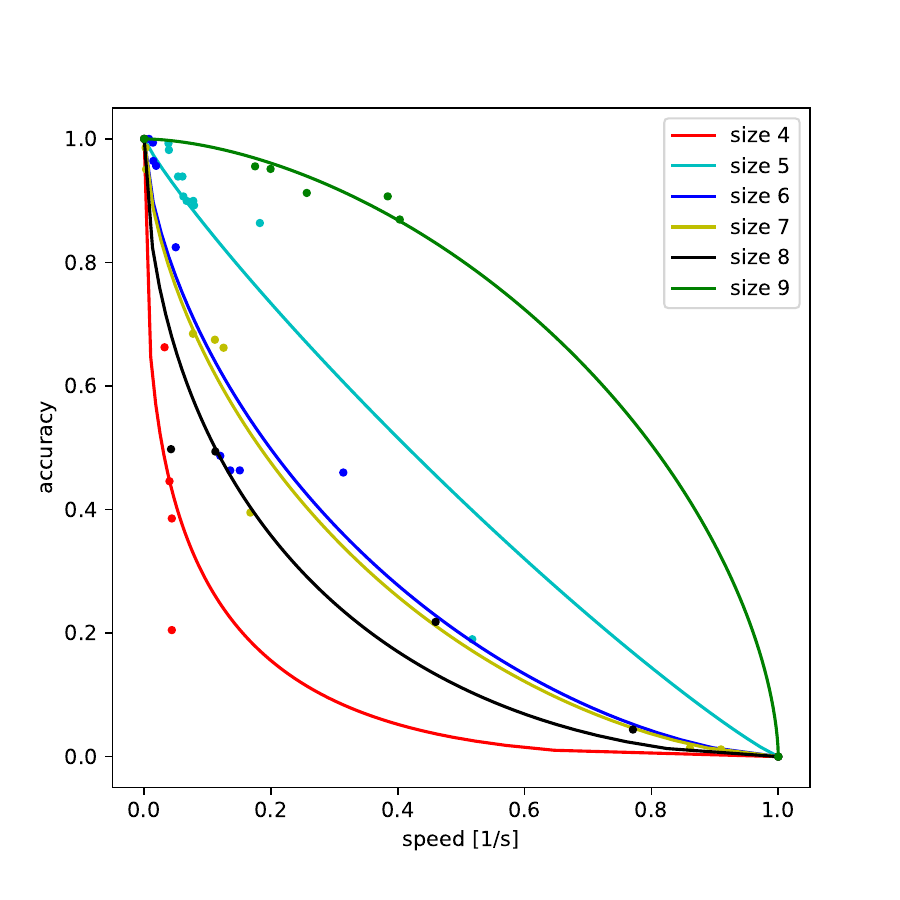}
        \caption{$L_p$ curves fitted to the Pareto fronts of multiple performance data. The accuracy and speed are normalized here.}
        \label{fig:fit_multi}
    \end{subfigure}
    \caption{Examples of $L_p$ curve fitting for quantum simulator performance data.}
\end{figure*}

 Scaling factors $a,b$ are taken in account in the area of the Lam\'e curve, but the offsets need to be included to calculate the full area. The offsets are included as:
 \begin{multline}
     A_{offset} = \min(\text{error})\cdot \max(\text{time}) + \\ \min(\text{time})\cdot \max(\text{error}) - \min(\text{error})\cdot \min(\text{time})
 \end{multline}
 Figure \ref{fig:area} shows a visual derivation of these areas.


The \textit{p}-norm in this context, indicates the stability of the solution for decreasing runtime. A low ($p<1$) norm indicates the accuracy drops steeply when decreasing runtime, whereas, with larger norms ($p\ge2$) the accuracy sustains and only drops at very small runtimes.

Since the curve expands by increasing accuracy, reducing time, and increasing the \textit{p}-norm, the partial score for size $n$ is defined as the surface $A_n$.
The total volume, and therefore the final score of the benchmark is then defined as the sum of the areas:
\begin{equation}
    \text{score} = \sum^n A_n
\end{equation}
With this definition, the score of a system will always increase if it allows for solving larger systems, but still scales with the performance. One drawback of the area metric is the possibility of including data points that are very fast but with very low accuracy. These data points will increase the area, without showing an increase in performance. To avoid this, a minimal accuracy requirement must be set. In this work, we have chosen a minimal accuracy requirement of 0.5.
\begin{center}
  \includegraphics[width=\linewidth]{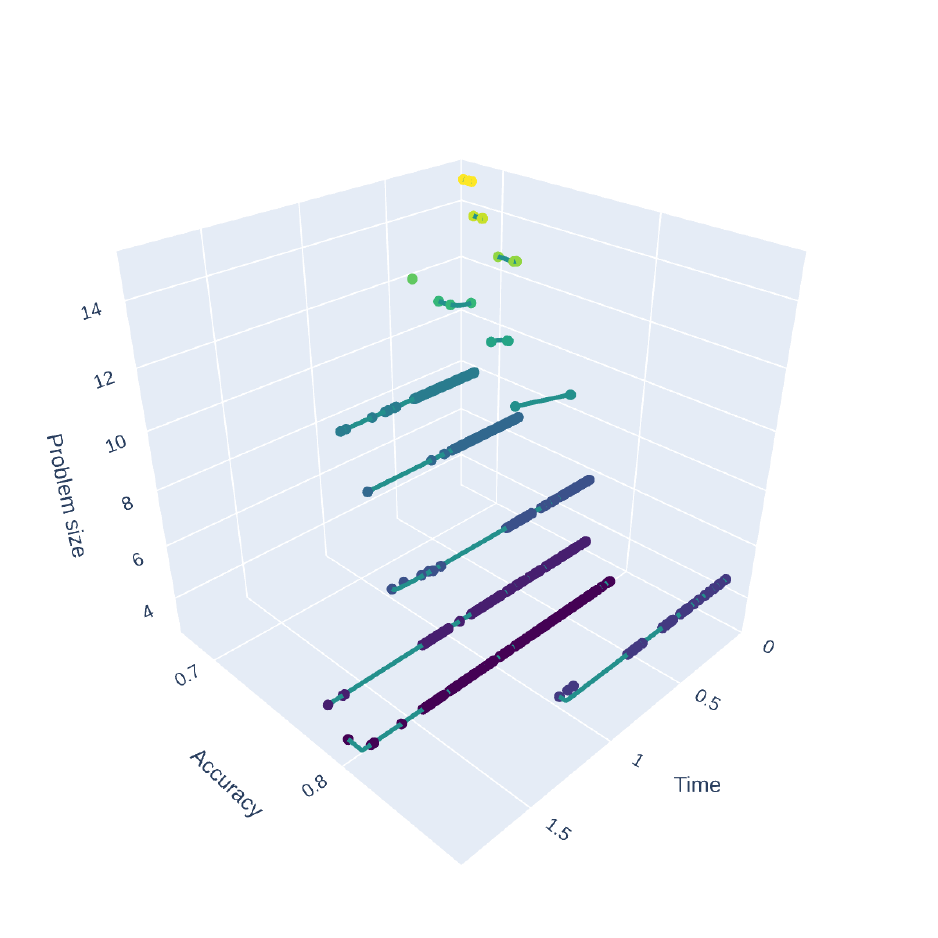}
  \captionof{figure}{3d plot of the fitted scores per problem size. Results are taken from the Qiskit-Aer simulator on a local CPU for problem sizes 3 to 15.}
  \label{fig:3d-score-sim}
\end{center}

Examples of the norm fitting are shown in Figure \ref{fig:normfit} and \ref{fig:fit_multi}. The data presented are results obtained using the IBM Aer simulator on a local CPU. A 3D plot of the fits for multiple problem sizes is shown in Figure \ref{fig:3d-score-sim}.

\section{Results}
\label{sec:results}
In this section, the QuAS metric will be evaluated using the KPIs and example problems mentioned above.

\begin{figure*}[t]
    \centering
    \begin{subfigure}{.4\linewidth}
    \centering
        \includegraphics[width=\linewidth]{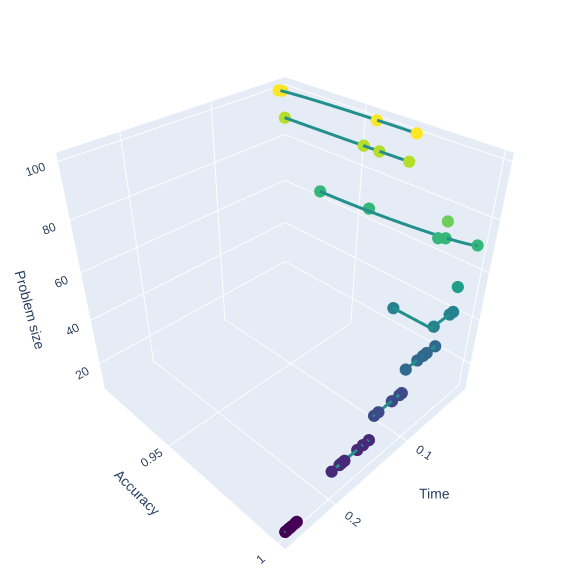}
        \caption{3D visualization of the Ising problem results on the D-Wave Annealer. Measurements are taken for problem sizes 10 to 100.}
        \label{fig:dwaveip}
    \end{subfigure}
    \begin{subfigure}{.4\linewidth}
    \centering
        \includegraphics[width=\linewidth]{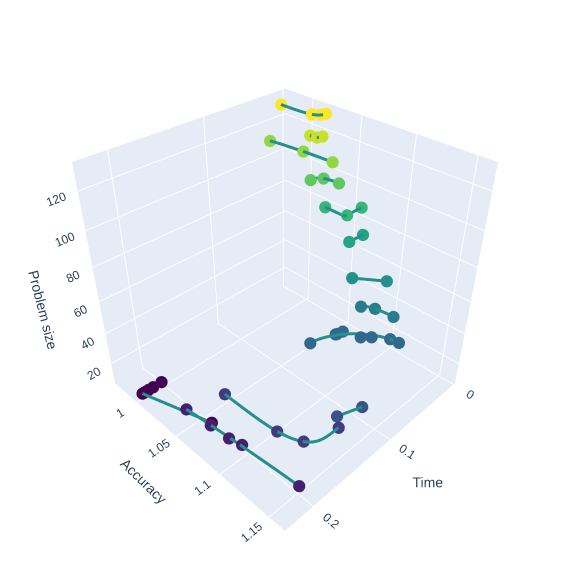}
        \caption{3D visualization of the max-cut problem results on the D-Wave Annealer. Measurements are taken for problem sizes 10 to 130.}
        \label{fig:dwavemcp}
    \end{subfigure}\\
    \begin{subfigure}{.4\linewidth}
    \centering
        \includegraphics[width=\linewidth]{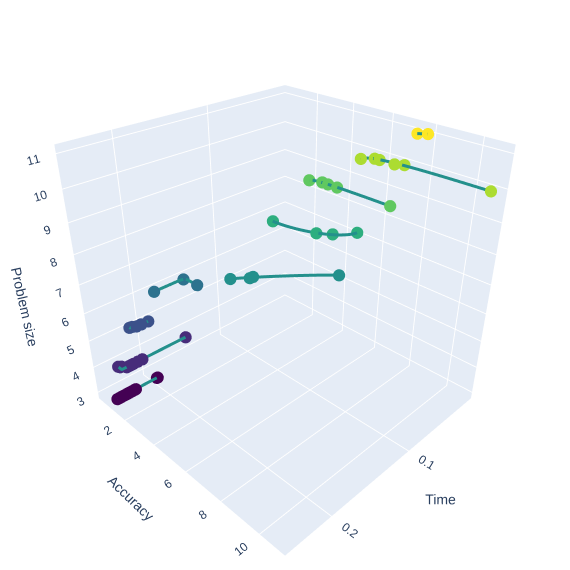}
        \caption{3D visualization of the traveling salesperson problem results on the D-Wave Annealer. Measurements are taken for problem sizes 3 to 11.}
        \label{fig:dwavetsp}
    \end{subfigure}
    \caption{$L_p$ curves fitted to measured accuracy and runtime performance data of the D-Wave Annealer and plotted in 3D for various problem sizes. The fitted curves are used to determine the sub-scores through the area calculation and aggregated to a single score.}
\end{figure*}

First, it is important to note that the definition of the metric is agnostic towards the type of processor or type of algorithms that can be used to solve the problem. This means that the metric can be used to benchmark any device that can solve the problem of interest, not just quantum devices. In addition, the metric is suitable to compare different algorithms run on the same hardware.

This implies that the metric can be used to benchmark quantum devices from a variety of quantum paradigms, such as gate-based quantum computing, quantum annealing and photonic quantum computing, as long as these quantum devices can solve the considered problem.

To show the broad capability of the metric, the following quantum solution methods will be used to benchmark the above problems: QAOA on (simulated) gate-based computing, and Quantum Annealing (QA). The QuAS score is capable of benchmarking classical solvers as well, but this is not explicitly demonstrated in this work.

To enable QuAS to run on different gate-based quantum backends, the tket library \cite{Sivarajah_2021} was used to handle circuit conversion and compilation. As such, the QuAS benchmark is applicable to a wide range of quantum vendors. In this work, a selection of the vendors is used for demonstration.
The gate-based algorithms will be solved using the Quantum Inspire simulator and Rigetti simulator. In addition, the metric will be evaluated using the D-Wave Advantage quantum annealer. The benchmark has also been run on IBMQ devices, but results have not been enough to construct a score for this device.

\begin{table}[h]
    \centering
    \begin{tabular}{|c|c|c|}
        \hline
         Vendor & Device & Max tested instance size \\
         \hline
         D-Wave & Advantage & 130\\
         \hline
         Quantum Inspire & QX simulator & 6 \\
         \hline
         Rigetti & PyQuil qvm & 20\\
         \hline
    \end{tabular}
    \caption{Specifics tested compute backend}
    \label{tab:backends}
\end{table}

For each combination of problem and solver, the problem will be run for 25 randomly generated instances for all problems. As a baseline, no scaling factors have been implemented.
Table~\ref{tab:backends} lists the specifics of each tested device.

The intermediate scores can be visualized by the curve fits as shown in Figures \ref{fig:dwaveip}, \ref{fig:dwavemcp} and \ref{fig:dwavetsp}, for the D-Wave device specifically. Here, measurements are shown for the IP, MCP and TSP for size 100, 130 and 11 respectively. For each problem size, the Pareto points are fitted to a Lame curve. The total score per instance is calculated as described in Section \ref{sec:method}. The results for the tested devices per instance are presented in Table \ref{tab:scores}. For these results, it should be noted that for the tsp problem, viable solutions were not always found. 
The score for the cost penalty for solutions not fitting the graph defaults to zero. This is caused by a restricted runtime on the devices. In this case, this does not allow for enough time for the optimization to find a viable result.

\begin{table}[h]
    \centering
    \begin{tabular}{|c|c|c|c|}
       \hline
       Device  & ip & mcp & tsp\\ 
       \hline
       D-Wave  & 1220.12 & 2082.06 & 66017.98 \\
       \hline
       Quantum Inspire & 493.52 & 0 & 0 \\
       \hline
       Rigetti qvm & 2.50 & 45.56 & 0 \\
       \hline
    \end{tabular}
    \caption{Score results per problem for tested backends.}
    \label{tab:scores}
\end{table}

\section{Conclusion and discussion}
\label{sec:conclusion}
To enable the full application of quantum devices, we need to be able to adequately measure the performance of these devices. This can be achieved by quantum metrics. There exists a wide variety of quantum metrics, each focusing on a different aspect of quantum computing. These metrics can be classified in three levels: component-level, system-level and application-level metrics. Application-level metrics are used to benchmark quantum devices by considering their performance on actual problems, of which many have already been proposed.

In this work, we define a new application-level metric called the Quantum Application Score (QuAS). This metric is defined by taking the strengths of the earlier metrics Q-score and QPack, together to form this new metric. The metric is designed as a framework to optimally benchmark (quantum) devices given a certain use case. Within this framework, various degrees of freedom can be tuned to match different use cases, namely the considered problem, the KPIs, and the weights of the KPIs. By tuning these parameters, the metric can be adjusted to different use cases.

The novelty of the QuAS comes from the fact that the KPIs and their weights can be changed freely. Earlier metrics have already considered the problem as a degree of freedom, but most currently available metrics output a single KPI in their metric score. Often, this KPI is either the accuracy or runtime. The QuAS is the first metric that allows flexibility between these KPIs. Looking at the potential end-use of quantum computing, this flexibility makes much more sense. For many use cases, it is not just about accuracy or runtime, but the correlation of these different KPIs when using a certain backend. The QuAS metric measures and quantifies this relationship between the different KPIs by computing a score that depends on these KPIs in an interdependent way. 
Even though there already exist various application-level metrics, we believe the novelty justifies the introduction of a new metric on this level. 

While this novelty forms a key strength of this metric, it has other characteristics that make it a suitable application-level metric, some of which we will discuss here. It should be noted that these are exactly the required characteristics for an application-level metric in section \ref{sec:related_work}. Firstly, the metric is easy to compare, as it consists of a single number. This also makes the metric easy to understand for quantum laymen. Secondly, the solver applies to any use case, as the problem to be considered can be freely chosen. Thirdly, it is \textbf{scalable}, as the metric does not require operations that become intractable for larger problem sizes. Lastly, the metric is defined in a completely \textbf{hardware-agnostic} way, meaning that any (quantum, classical, or other) solver capable of solving the problem at hand, can be benchmarked using the metric. 
This characteristic is often absent in alternative benchmarks, which focus mainly on gate-based quantum computers. 


To showcase the applicability of the QuAS metric, it has been used to benchmark quantum annealers, as well as gate-based quantum simulators. The results of these benchmarks show a significant score advantage for the D-Wave Annealer compared to the tested gate-based quantum simulators. The reason for the difference is the limited number of qubits that can be simulated using a local simulator, as well as the significant time it takes to simulate the circuits. As the runtimes are limited, the optimization will rarely find the optimal results. This is most notable in the Quantum Inspire measurement for the max-cut problem, as the optimizer did not find optimization results within a reasonable time. Similarly, measurements taken on the IBM quantum hardware took a significant amount of iterations, resulting in too few measurements for constructing a score.

While the QuAS metric is applicable in many situations, there are a couple of requirements these situations need to adhere to. First and foremost, the QuAS metric requires the problem to be formulated as an optimization problem. Specifically, the problem is required to have its solutions expressible as a single value, with a higher value indicating a better performance (i.e., single-objective optimization). While this is true for many problems, such as the one posed in this work, Problems for which multiple objectives need to be optimized do not apply to QuAS. This limits the applicability of the metric.


A challenge with defining the KPIs is that measured values are not bound. With the progression of quantum computers, solvable problem sizes increase, runtimes get shorter and accuracies get better. While accuracy is bound by some extent (the relative difference between heuristic and best solution), the other KPIs are not. This can result in the metric scores increasing indefinitely. While this scaling is inevitable, we believe that the current scaling of the KPIs still makes for a fair comparison of solver backends and will scale well for future devices.

Another challenge in designing quantum benchmarks is how the metric will scale with error correction. M. Amico et al. \cite{amico2023defining} argue that error correction will give an unfair advantage if the metric does not reflect the overhead in the performance measure. Error correction will result in relatively higher accuracy but longer runtime. In our work, we argue that QuAS reflects this tradeoff accurately. As QuAS combines both accuracy and runtime into a single surface, the combination gives a fair reflection of the quantum hardware performance.

The QuAS metric has an inherent focus on the size KPI, as the metric is defined as the sum of different problem sizes. While this doesn't impede the definition of the QuAS metric, it does put an unbalanced focus on this KPI. This is not completely in the spirit of the QuAS metric, as the QuAS metric is defined so that the weights between these different KPIs can be chosen freely. There are many use cases for which the problem size is less relevant or not even relevant at all, for which the current definition of QuAS does not allow. This issue could be circumvented by redefining the QUAS as follows. Again, each data point would be represented in 3D space (or $n$-D if the QuAS is instantiated with $n$ KPIs), for which the Pareto front could directly be constructed. This three-dimensional Pareto front could then be fitted to a three-dimensional Lamé curve, after which the QuAS can be defined as the volume of the shape bounded by this curve. This raises some issues, however. Most notably, whether a suitable Pareto front could be defined in 3D space, and whether the Lamé curve fitting works well in 3D. In addition, this requires the size metric to lie within a given interval, which will not remain true with the increasing problem-solving abilities of (quantum) computers. Further research will have to determine if a redefinition of the QuAS as such would work.

It is important to mention that the QuAS doesn't define a metric, but defines a framework for benchmarking (quantum) devices. For each choice of problem and KPIs, the QuAS framework yields a new metric. Because of this, it is more difficult to compare devices using a QuAS metric, as the benchmark needs to be specific about the choices that were made for the QuAS instantiation. This could be seen as a weakness of the QuAS metric. However, we believe that to optimally benchmark devices at an application level, a framework is more suitable than a single metric. Only with a framework in which certain characteristics can be changed dynamically, the specific nuances of a single use case can be optimally benchmarked.

In this work, we have shown that the QuAS metric can be implemented on quantum annealers as well as simulated gate-based quantum devices. We have also implemented the metric on IBM hardware, but unfortunately due to budget limitations, could not produce sufficient results to deduce a QuAS for this hardware. Although this is a shame, of course, it does not weaken the position of this work. This work aims to define a new metric and showcase its applicability to various quantum devices. By running experiments on (simulated) quantum hardware, we have demonstrated QuAS applicability and its capabilities.

Budget constraint issues also tie in with another important factor in quantum benchmarking. To reach the full potential of quantum metrics, it is important that the industry, and specifically quantum hardware providers, start using these metrics to benchmark their own devices. Specifically, these metrics should then be used to steer the development of these devices. To achieve this, hardware providers need to be supportive of implementing these metrics on their devices. This would allow for demonstrating new metrics on varied backends, giving a more balanced and fair display of metric utility. Without vendor support, quantum metrics are limited in improving quantum devices.

For further research, there are a couple of avenues to explore. Firstly, it would be interesting to apply other instantiations of the QuAS metric, specifically with other KPIs, other weights between the KPIs, and with other problems, to showcase that the QuAS metric has broad applicability to various use cases. In addition, it would be interesting to explore whether the redefinition of the QuAS as above would be feasible. Secondly, it would be interesting to apply the QuAS to other hardware types and computer paradigms, to further showcase the applicability of the metric to different solver types. Lastly, it would be interesting to 'benchmark the benchmarks': this could for example be done by using the QuAS and other quantum metrics to perform the same benchmark. In this way, currently available quantum metrics could be compared, showing the strengths and weaknesses of each of them.

\bibliographystyle{IEEEtran}
\bibliography{bib}

\end{document}